\documentclass[aps,showpacs]{revtex4}

\usepackage{graphicx}
\usepackage{amsmath}
\usepackage{amssymb}

\begin{document}

\def\a{\alpha}
\def\b{\beta}
\def\e{\varepsilon}
\def\d{\delta}
\def\m{\mu}
\def\t{\tau}
\def\n{\nu}
\def\o{\omega}
\def\s{\sigma}
\def\G{\Gamma}
\def\D{\Delta}
\def\O{\Omega}

\def\ra{\rightarrow}
\def\Ra{\Rightarrow}
\def\pd{\partial}
\def\bk{{\bf k}}
\def\bq{{\bf q}}
\def\bQ{{\bf Q}}

\def\be{\begin{equation}}\def\ee{\end{equation}}
\def\bea{\begin{eqnarray}}\def\eea{\end{eqnarray}}
\def\nn{\nonumber}
\def\lb{\label}
\def\pref#1{(\ref{#1})}


\title{Effect of orbital currents on the restricted optical
conductivity sum rule}

\author{L.~Benfatto$^{1}$}
\author{S.G.~Sharapov$^{2}$}
\thanks{Present address: Institute for Scientific Interchange, via Settimio
Severo 65, I-10133 Torino, Italy}
\author{H.~Beck$^2$}%
\affiliation{$^1$D\'epartment de Physique,
        Universit\'e de Fribourg, P\'erolles, CH-1700 Fribourg, Switzerland\\
        $^2$Institut de Physique,
        Universit\'e de Neuch\^atel, CH-2000 Neuch\^atel, Switzerland}

\date{\today }

\begin{abstract}
We derive the restricted optical-conductivity sum rule for a model with
circulating orbital currents.  It is shown that an unusual coupling of the
vector potential to the interaction term of the model Hamiltonian results
in a non-standard form of the sum rule. As a consequence, the temperature
dependence of the restricted spectral weight could be compatible with
existing experimental data for high-$T_c$ cuprates above
the critical temperature $T_c$.  We extend our results
to the superconducting state, and comment on the differences and analogies
between these two symmetry-breaking phenomena.
\end{abstract}

\pacs{71.10.-w, 74.25.Gz, 72.15.-v, 74.72.-h}



\maketitle

\section{Introduction}

Recent results extracted from measurements
\cite{Molegraaf:2002:Science,Sanatander:2002:PRL,Homes:2004:PRB,
Santander:2004,Boris:2004:Science} of the in-plane optical conductivity
$\sigma(\omega)$ in Bi$_2$Sr$_2$CaCu$_2$O$_{8+\delta}$ (BSCCO) place strong
constraints on possible theories of high-$T_c$ superconductivity.  Special
attention has been devoted to the partial spectral weight extracted from
the optical conductivity in the direction $\a=x,y$
\begin{equation}
\label{weight}
W_\a(\omega_m,T) = \int_{-\omega_m}^{\omega_m} \mbox{Re} \,
\sigma_{\a\a}(\omega,T)d \omega,
\end{equation}
which is analyzed as a function of temperature $T$
and the cutoff frequency $\omega_m$, which varies between $1000$ cm$^{-1}$
($0.12$ eV) and $20 000$ cm$^{-1}$ ($2.5$ eV).
According to this definition, the weight $W$ includes  the
condensate peak at $\omega=0$ which develops in the superconducting (SC) state
below $T_c$. When $\omega_m$ is of order of the plasma frequency,
$\o_P\varpropto 10^4$ cm$^{-1}$,  only intraband optical
transitions contribute to the measured spectral weight (\ref{weight}), and
the so-called {\em restricted} or {\em partial sum rule}
may be applied \cite{Maldague:1977:PRB,Baeriswyl:1987:PRB,Scalapino:1993:PRB},
which relates $W$ to the average value of the diamagnetic term $\t_{\a\a}$
(see Eq.~(\ref{optical.conductivity}) below),
\bea
W_\a(\o_P,T)\equiv W(T) =
\frac{\pi e^2}{V} \langle \tau_{\a\a} \rangle=\nn\\
\frac{\pi e^2}{V N} \sum_{\mathbf{k},\sigma}
\frac{\partial^2 \e_\bk }{\partial k_\a^2} n_{\bk,\sigma}
= - \frac{\pi e^2}{V} \frac{\langle K \rangle}{d},
\label{weight.band}
\eea where $n_{\bk,\sigma}$ is the momentum occupation number, $V$ is the
unit-cell volume, $N$ is the system size, $d=2$ is the dimension of
the system, $e$ is the electron charge, and we set $\hbar = c =1$.  The
second line of Eq.~(\ref{weight.band}) is obtained under the assumption
that the interaction term of the Hamiltonian {\em does not couple to the
vector potential $\mathbf{A}$\/}, and the final equality in
Eq.~(\ref{weight.band}) is valid only for a nearest-neighbors tight-binding
dispersion $\e_\bk = - 2t (\cos k_x a + \cos k_y a)$ with the lattice
constant $a$.  In this case the spectral weight is a direct measure of the
mean kinetic energy $K$ of the system, and depends on temperature and
interaction strength. For a non-interacting system
$n_{{\bf k} \sigma} =f(\xi_\bk)$, where $\xi_\bk=\e_\bk-\mu$, $\mu$ is the
chemical potential, and $f(x)$ is the Fermi function, so that
$W(T)$ increases as
the temperature decreases. In the presence of a SC instability, the BCS
theory predicts that the occupation number is modified below $T_c$ and the
partial spectral weight is
\begin{equation}
\label{T.SC}
\frac{W(T) }{\pi e^2 a^2}=- \frac{1}{2 V N}\sum_{\mathbf{k}} \e_\bk
\left[1 - \frac{\xi_\bk}{E_\bk^{SC}}
\tanh \frac{E^{SC}_\bk}{2T} \right]
\end{equation}
where $\Delta_\bk =(\Delta_0/2) (\cos k_x a - \cos k_y a)$ is the $d$-wave
SC gap and $E^{SC}_\bk = \sqrt{\xi^2_\bk + \Delta^2_\bk}$ is
the quasiparticle dispersion in the SC state.  The kinetic energy
increases below $T_c$ because of particle-hole mixing, and as a
consequence $W(T)$ decreases. The restricted sum rule should be contrasted
\cite{Jaklic:2000:AP,Millis:book,Marel:2003,Marel:2003a} with the full f-sum rule
\begin{equation}
\label{full.rule}
\int_{-\infty}^{\infty} \mbox{Re} \, \sigma(\omega) d \omega
= \frac{\pi n e^2}{m},
\end{equation}
which relates the integral over all optical transitions
($\omega_m \to \infty$) to the total carrier density $n$
and bare mass $m$, and is independent of temperature and interactions.
The difference between the full and restricted sum rules is made up
by transitions between the orbitals described by the low-energy
effective tight-binding model and orbitals
with the energies above $\omega_P$, not included in this model.
As noted in Ref.~\cite{Millis:book}, there is as yet no complete
understanding of the relevant orbitals or energy range
over which the full sum rule is restored.

The experimental results and their implications
\cite{Molegraaf:2002:Science,Sanatander:2002:PRL,Santander:2004}
for the restricted sum rule  are the following. (i) Above the
critical temperature $T_c$, the partial spectral weight $W(T)$
{\em does not decrease\/} when $T$ {\em decreases\/}. Assuming
that the mean-field Eq.~\pref{T.SC} is already valid in the
pseudogap state for $T >T_c$, so that $W(T)$ would decrease as $T$
decreases even above $T_c$, following
Ref.~\cite{Sanatander:2002:PRL} one may conclude  that the
observed increase of $W(T)$ is in contradiction with the opening
of a pseudogap. (ii) Using the value $t=0.25$ eV, which is typical
for cuprates,  the tight-binding estimate Eq.~\pref{weight.band}
for the relative thermal variation of $W(T)$ between $T_c$ and the
room temperature $T_r$, with $T_r/t\simeq 0.1$, gives
$(W(T_c)-W(T_r))/W(T_r)\simeq 2 \cdot 10^{-3}$. This variation
appears to be at least one order of magnitude smaller than the
experimentally observed change in $W(T)$ even for a large $\omega_m =
10^4 \mbox{cm}^{-1}$. There is even faster increase of $W(T)$
when smaller values of $\omega_m$ are used \cite{Santander:2004}. 
(iii) Below $T_c$ the situation is not clear yet.
While early measurements in BSCCO samples show that
there is an even faster {\em increase\/} of $W(T)$
\cite{Molegraaf:2002:Science, Sanatander:2002:PRL}, contrary to
the prediction of the BCS theory, more recent results in BSCCO 
\cite{Santander:2004} show that there is a flattening
of $W(T)$ in underdoped samples for $\omega_m = 8000
\mbox{cm}^{-1}$, while a BCS behavior below $T_c$ is seen in the overdoped
BSCCO and in YBCO samples \cite{Homes:2004:PRB, Boris:2004:Science}.

The possibility of a spectral-weight change below the superconducting
critical temperature has been analyzed, for example, in 
Refs.~\cite{Hirsch:2000:PRB,Norman:2002:PRB,Eckl:2002} in terms of the 
lowering of the in-plane kinetic energy. 
In Ref.~\cite{Eckl:2002}) the reduction of 
the kinetic energy at $T_c$ has been attributed to the transition from a 
phase-incoherent Cooper pair motion in the pseudogap regime above $T_c$ 
(see, e.g. review \cite{Loktev:2001:PRP})
to a phase coherent motion at $T_c$, while in 
Ref.~\cite{Norman:2002:PRB}
a model with a frequency dependent scattering rate was used.
More recently, the
optical conductivity sum rule has been analyzed for a model
with electron coupled to a single Einstein oscillator  \cite{Knigavko:2004}. 

Here, in contrast to these papers, we focus primarily on the temperature 
dependence of $W(T)$ above $T_c$ and on the issue of the compatibility
between the pseudogap opening and the absence of a lowering of the 
spectral weight. Our purpose is to show that
if the pseudogap originates from a state with circulating orbital currents
\cite{Halperin:1968:SSP,Affleck:1988:PRB,Nersesyan:1989:JLTP,
Schulz:1989:PRB,Varma:1997:PRB,Eremin:1998:JETPL,Benfatto:epj,
Chakravarty:2001:PRB}
the opening of the pseudogap can be accompanied by the
increase of the partial spectral weight. For completeness we extend these
results also to the SC state. We find that when a small SC gap opens in the
presence of a large DDW gap $W(T)$ remains almost constant below $T_c$.

\section{Model}
We consider the
model with bond currents circulating around elementary
plaquettes of copper atoms which is described by the effective Hamiltonian
\begin{equation}
\label{Hamiltonian}
H =  \sum_{\mathbf{k},\sigma} [\xi_\bk
c_{\bk\sigma}^\dagger c_{\bk\sigma} +
i D_\bk c_{\bk\sigma}^\dagger c_{\bk+\bQ\sigma}],
\end{equation}
where $c_{\bk\sigma}^{\dagger}$, $c_{\bk\sigma}$ are creation and
annihilation operators for a particle with momentum $\mathbf{k}$ and spin
$\sigma$, $D_\bk = (D_0/{2})(\cos k_x a -\cos k_y a )$ is the gap, known as
the DDW gap \cite{Chakravarty:2001:PRB}, arising from the formation of the
state with circulating currents, and $\mathbf{Q} = (\pi/a, \pi/ a)$ is the
wave vector at which the density-wave ordering takes place. In the present
paper we do not derive the Hamiltonian \pref{Hamiltonian} by means of
a Hartree-Fock analysis of a microscopic model, as it has been done
elsewhere \cite{Halperin:1968:SSP,Affleck:1988:PRB,Nersesyan:1989:JLTP,
Schulz:1989:PRB,Varma:1997:PRB,Eremin:1998:JETPL,Cappelluti}.
These studies showed that Hubbard-like
Hamiltonians with additional finite-range repulsion and
superexchange interaction can have a stable DDW saddle
point. Thus we shall parametrize phenomenologically the DDW order parameter
and we will analyze within the low-energy effective model
\pref{Hamiltonian} the effect of this symmetry breaking on the optical sum
rule. It is worth noting that this approach has been often adopted in the
literature to address several issues related to transport properties in the
DDW state \cite{Benfatto:epj,Chakravarty:2001:PRB,
Tewari:2001:PRB,Wang:2001:PRL,Yang:2002:PRB,Kee:2002:PRB,morr,
Chakravarty:2002:PRL,Sharapov:2003:PRB}.

The derivation of the restricted sum rule depends crucially on the manner
in which the vector potential $\mathbf{A}$ enters the effective low-energy
Hamiltonian~(\ref{Hamiltonian}). For lattice models $\mathbf{A}$ is usually
inserted in a coordinate representation by means of the {\em Peierls
ansatz} \cite{Jaklic:2000:AP,Millis:book,Scalapino:1993:PRB} $c_i\ra c_i
e^{- i e\int {\mathbf A}\cdot d {\mathbf r}}$, which modifies the fermionic
operator at every site $i$. The dependence of the resulting Hamiltonian on
each component $A_\a$ of the gauge field is $ H(A_\a) \approx H(0) -\sum_i
\left[ e A_\a(i) j^P_\a(i) - \frac{e^2}{2} A_\a^2(i) \tau_{\a\a}(i)
\right], $ where $j^P_\a(i)$ is the $\a$ component of the particle current
density and $\tau_{\a\a}(i)$ is the $\a\a$ component of the diamagnetic
contribution.  Thus the total current density $j_\a(i)$ is $ j_\a (i)= -
\delta H/\delta A_\a(i) = e j^P_\a - e^2 \tau_{\a\a}(i) A_\a(i).$ By
evaluating $\langle j_\a (\omega) \rangle$ in linear response
\cite{Scalapino:1993:PRB,Enz:book}, one obtains the complex optical
conductivity
\bea
\sigma_{\a\a} (\omega) &=& \frac{i e^2}{V(\omega + i 0)}{\cal K}_{\a\a}(\o,{\bf
0})\nonumber\\
&=&\frac{i e^2}{V(\omega + i 0)}
(\langle \tau_{\a\a} \rangle - \Lambda_{\a\a}(\omega, \mathbf{0})),
\label{optical.conductivity}
\eea
where ${\cal K}_{\a\a}$ is the electromagnetic response kernel and the
current-current correlation function
$\Lambda_{\a\a}(\omega, \mathbf{q})$ is defined as
\begin{equation}
\Lambda_{\a\a}(i \Omega_n, \mathbf{q}) = \frac{1}{N}\int_0^\beta d \tau
e^{i \Omega_n \tau}
\langle j^P_\a(\tau, \mathbf{q}) j^P_\a(0, - \mathbf{q}) \rangle,
\end{equation}
with $\Omega_n = 2 \pi n T$, using the standard analytic continuation
$i \Omega_n \to \omega + i0$.
Here $j_\a^P(\tau, \mathbf{q})$ is the Fourier transform of the
current density expressed in imaginary-time representation.
The Kramers-Kronig relations for the response function
$\Lambda_{\a\a}(\omega, \mathbf{0})$, yield from
Eq.~(\ref{optical.conductivity}) the optical sum rule \pref{weight.band}
for the tight-binding model with nearest-neighbors hopping.
This derivation of the optical-conductivity sum rule requires the
knowledge of $\tau_{\a\a}$, which is easily obtained
for a  Hamiltonian expressed in coordinate representation. For a
Hamiltonian in the momentum representation,
it is more straightforward to apply the sum rule in the form \cite{Enz:book}
\begin{equation}
\label{sum.momentum}
\int_{- \o_P}^{\o_P} \mbox{Re} \sigma (\omega) d \omega =
\frac{\pi e^2}{V N} \lim_{q_\alpha \to 0} \frac{1}{q_\alpha}
\langle[\rho(t,\mathbf{q}), j^{P}_\alpha(t,-\mathbf{q})] \rangle,
\end{equation}
where $\rho(t,\mathbf{q})$ is the Fourier transform
of the particle density which satisfies the continuity equation
\begin{equation}
\label{charge.conservation}
\frac{\partial \rho(t, \mathbf{q})}{\partial t} + i \mathbf{q}
\cdot \mathbf{j}^P(t, \mathbf{q}) = 0.
\end{equation}
We note that substitution in Eq.~(\ref{sum.momentum})
of $\rho(t,\mathbf{q}) = \sum_{\mathbf{k}, \sigma}
c^{\dagger}_{\mathbf{k} - \mathbf{q}/2,\s}
c_{\mathbf{k}+ \mathbf{q}/2,\s}$
and of  the free-electron expression $j^{P}(t, \mathbf{q}) =
(1/m) \sum_{\mathbf{k}, \sigma} \mathbf{k}
c^{\dagger}_{\mathbf{k} - \mathbf{q}/2,\s}
c_{\mathbf{k}+ \mathbf{q}/2,\s}$, corresponding to $\e_\bk=\bk^2/2m$,
returns the full f-sum rule (\ref{full.rule})  (see \cite{Enz:book}).

In most cases it is assumed that the interaction term of
the Hamiltonian involves only density-density coupling, so that
this is trivially gauge invariant and the Peierls ansatz modifies
only the first, ``kinetic'', term of the Hamiltonian
(\ref{Hamiltonian}). For models with nearest-neighbors hopping,
$\tau_{\a\a}$ is then related directly to the kinetic energy,
$\langle \tau_{\a \a} \rangle = -\langle K \rangle/d$
and one obtains the usual version of the sum rule
given by Eq.~(\ref{weight.band}). However, this assumption is invalid when
for example ``occupation modulated'' hopping terms are
considered \cite{Hirsch:2000:PRB}.  In particular, if one assumes that the
low-energy physics of the system can be described by the effective
Hamiltonian (\ref{Hamiltonian}), then any distinction
in the total energy between a kinetic and a potential part is somehow
ambiguous. Thus, by transforming the Hamiltonian (\ref{Hamiltonian}) to
the coordinate space, one finds that $\tau_{\a\a}$
contains an extra term for $D_0\neq 0$,
\be
\langle \t_{\a\a}\rangle=-\frac{1}{2N}\sum_{\bk\s}\e_\bk \langle
c_{\bk\s}^\dagger c_{\bk\s} \rangle+
i D_\bk \langle c_{\bk\s}^\dagger c_{\bk+\bQ\s}\rangle.
\label{new.tensor}
\ee
This
result is consistent with the derivation \pref{sum.momentum} of the sum
rule, when one uses the particle current operator compatible with the
conservation law (\ref{charge.conservation}) and with
the equations of motion for the operators $c$ and $c^\dagger$,
\cite{Tewari:2001:PRB,Yang:2002:PRB,Kee:2002:PRB,Sharapov:2003:PRB}
\begin{equation}
\label{electric.current.DDW}
\mathbf{j}^P (t, \mathbf{q})=\sum_{\mathbf{k}, \sigma}
\left[ v_\bk^F c^{\dagger}_{\mathbf{k}- \mathbf{q}/2\s}
c_{\mathbf{k}+\mathbf{q}/2\s}
-iv_\bk^D c^{\dagger}_{\mathbf{k}- \mathbf{q}/2\s}
c_{\mathbf{k} + \mathbf{Q}+ \mathbf{q}/2\s} \right],
\end{equation}
where $v_\bk^F=\partial \e_\bk/\partial \mathbf{k}$ and $v_\bk^D=-\partial
D_\bk/\partial \mathbf{k}$. The first term of the previous expression
relates as usual the particle current to the band velocity $v_\bk^F$. The
second term, which only appears for non-vanishing $D_0$, takes into account
the contribution of the orbital currents to the electrical conductivity,
arising when the DDW order is established.  Substitution of
Eq.~(\ref{electric.current.DDW}) in (\ref{sum.momentum}) yields
\begin{equation}
\label{sum.T>Tc}
\frac{W(T)}{\pi e^2a^2}=-
\frac{1}{V N}\sum_{\mathrm{RBZ}}
E_\mathbf{k}
[f (\xi_{+,\mathbf{k}}) - f(\xi_{-,\mathbf{k}})],
\end{equation}
where $E_\mathbf{k}=\sqrt{\e^2_\mathbf{k}+ D^2_\mathbf{k}}$, and
$\xi_{\pm,\mathbf{k}} = - \mu \pm E_\mathbf{k}$ represent the two
excitation branches associated with the formation of DDW order which breaks
translation symmetry.  The sum is taken over the reduced Brillouin zone
(RBZ).  Eq.~(\ref{sum.T>Tc}) was derived using the fact that
 $\partial_\alpha v^F_\bk= 2ta^2 \cos k_\a a$ (and
$\partial_{x,y} v^D_\bk= \pm (D_0/2) a^2 \cos k_{x,y}a$), and it reduces to
Eq.~(\ref{weight.band}) for $D_0=0$.

To extend this result to a state with both DDW and SC order present, we
add to the Hamiltonian (\ref{Hamiltonian}) an additional d-wave
mean-field pairing
term $H_p= \sum_{\mathbf{k},\sigma} [\Delta^{\ast}_\bk
c_{-\bk\downarrow}c_{\bk\uparrow} + h.c.]$, where $\Delta_\bk=(\Delta_0/2)
(\cos k_x a-\cos k_y a)$.
As a consequence, the spectral weight in the DDW+SC state reads
\begin{equation}
\label{sum.complete}
\frac{W(T)}{\pi e^2 a^2}=
\frac{1}{2V N}\sum_{\mathrm{RBZ}} E
\left[\frac{\xi_{+}}{E_+} \tanh \frac{E_+}{2T} -
\frac{\xi_{-}}{E_-}
\tanh \frac{E_-}{2T} \right],
\ee
where
$E_{\pm,\mathbf{k}}=\sqrt{\xi_{\pm,\bk}^2+\Delta^2_\mathbf{k}}$
is the quasiparticle dispersion in the presence of pairing, and the explicit
dependence on $\bk$ has been omitted.
\begin{figure}
\centering{
\includegraphics[width=6.cm, angle=-90]{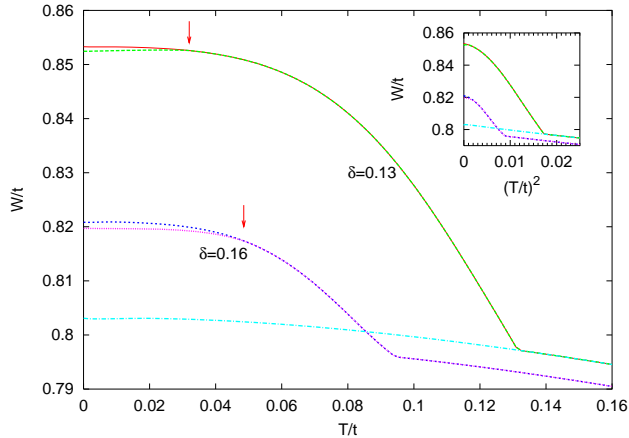}
}
\caption{$W(T)/t$ in units of $e^2\pi a^2/V$ [Eq.~(\ref{sum.T>Tc})]
for an underdoped ($\delta=0.13$) and an optimally
doped ($\delta=0.16$) system. Below $T_c$, marked by the arrows, the
lower line for each doping represents the spectral weight in the DDW+SC state,
Eq.~(\ref{sum.complete}). We note that for $\d=0.13$ the decrease of
$W(T)$ below $T_c$ is almost negligible. Also shown for
comparison (dash-dotted line) is $W(T)/t$ in the normal state without DDW
formation at $\d=0.13$ [Eq.~(\ref{weight.band})].
Inset: spectral weight plotted as  function of $(T/t)^2$.}
\label{fig:1}
\end{figure}
Numerical calculation of the spectral weight defined by
Eqs.~\pref{sum.T>Tc} and \pref{sum.complete}  shows
that below the temperature $T_{DDW}$, at which the DDW state
is formed, the spectral weight $W(T)$ increases as
the temperature decreases and the DDW gap opens. As stated
above, this increase originates from the second term of
Hamiltonian~\pref{Hamiltonian} which effectively enhances the low-frequency
conductivity, as already observed in Ref.~\cite{Sharapov:2003:PRB}.
When the temperature is lowered further and the SC gap opens
at a temperature $T_c<T_{DDW}$, the spectral-weight increase is reduced
with respect to the DDW state only.

A more quantitative comparison with experimental data requires the
dependences of the gaps $D_0$ and $\Delta_0$ on temperature $T$
and doping $\delta$. This issue has been investigated within
various microscopic models by several authors (see for example
\cite{Eremin:1998:JETPL,Cappelluti}). Here, following the lines of
Ref.~\cite{Benfatto:epj,Tewari:2001:PRB,Chakravarty:2002:PRL}, and
consistently with our effective Hamiltonian (\ref{Hamiltonian}),
we adopt a mean-field dependence for the DDW gap. We assume
\cite{Benfatto:epj} that $D_0$ opens below a doping-dependent
temperature $T_{DDW}(\d)=40[1-(\delta/\delta_0)^4]$ meV, where
$\delta$ is the doping with respect to half filling, and
$\delta_0=0.2$ is the critical doping for the DDW formation. We
adopt for the temperature dependence of $D_0$ the mean-field
relation $D_0(T,\d)=c T_{DDW}(\d)g(T/T_{DDW}(\d)), $ where
$g(x)=(1-x^4/3)\sqrt{1-x^4}$, and $c$ is a constant which is used
as a fitting parameter. To describe the SC transition we solve
self-consistently the BCS equations for $\D_0$ and $\mu$ as
functions of temperature. We use an analogous set of parameters as
in Ref.~\cite{Benfatto:epj} to estimate the thermal variation of
$W(T)$ in BSCCO. We focus on an underdoped ($\d=0.13$) and an
optimally doped ($\d=0.16$) compound
\cite{Molegraaf:2002:Science,Sanatander:2002:PRL}. The results are
presented in Fig.\ 1, where the temperature dependence of $W(T)$
for the tight-binding metal is shown for comparison. Below
$T_{DDW}$ the weight $W(T)$ increases proportionally to $D_0(T)$,
so that the overall increase of $W(T)$ with respect to
$W(T_{DDW})$ is more pronounced in the underdoped case
($\d=0.13$), where $D_0(0)$ is larger. In
Ref.~\cite{Molegraaf:2002:Science} it has been observed that
$W(T)$ shows a $T^2$ dependence above the critical temperature, as
expected in the tight-binding model Eq.~(\ref{weight.band}), but
with a much larger slope. To make a comparison with the result of
Eq.~(\ref{sum.T>Tc}), in the inset of Fig.\ 1 we plot $W(T)$ as a
function of $T^2$.  One can see that below $T_{DDW}$ the $T^2$
temperature dependence of $W(T)$ is still recovered over a wide
range of temperature, and with a slope in good agreement with the
experimental observation. Significant deviations are observed
approaching $T_c$ and below, where our mean-field approach does
not reproduce the anomalous increasing of $W(T)$ observed in BSCCO
in early experiments \cite{Molegraaf:2002:Science}. Observe
however that because at these dopings $\D_0\ll D_0$, $W(T)$ is not
explicitly decreasing below $T_c$, as expected in an ordinary
metal-SC transition, but keeps almost constant resembling more
recent experimental data \cite{Santander:2004}. Finally, we find
that the relative variations of $W(T)$ below $T_r=0.1t$ is of
order  $W(0)/W(T_r)\simeq 3 \cdot 10^{-2}$, as observed
experimentally
\cite{Molegraaf:2002:Science,Sanatander:2002:PRL,Santander:2004},
and is much larger than expected in the simple tight-binding
model.

The previous analysis can be extended to the case where
an additional next-nearest neighbor hopping term $t'$ is
added to the bare band dispersion $\xi_\bk$ in
Eq.~(\ref{Hamiltonian}). Even though both Eq.~(\ref{new.tensor}) and
Eq.~(\ref{sum.T>Tc}) are formally modified, the qualitative behavior of the
reduced spectral weight is the same,
with an increasing of $W(T)$ below the temperature for DDW formation.
However, as suggested
also in Ref. \cite{morr}, it is likely that the analysis of the DDW state
should be carried out with a value of $t'$ much smaller then suggested by
ARPES experiments, leading to small quantitative corrections to the
previous results. Also,
as shown in \cite{Marel:2003a}, the effect of including
next-neighbors hopping term $t^{\prime }$ is small, so for qualitative
considerations one may consider the model only with a hopping $t$.

\section{Discussion}

A crucial step in the presented derivation of the sum rule is to use the
current operator (\ref{electric.current.DDW}) that was considered before in
Refs.~\cite{Tewari:2001:PRB,Yang:2002:PRB,Kee:2002:PRB,Sharapov:2003:PRB}.
A different current operator was used instead in
Ref.~\cite{Chakravarty:2002:PRL}, where it was suggested that the gauge
field should couple via the Peierls ansatz to the quasiparticle fermionic
operators that diagonalize the Hamiltonian (\ref{Hamiltonian}).  As far as
the restricted sum rule is concerned, this corresponds to the replacement
of the bare dispersion law $\e_\bk$ in Eq.~(\ref{weight.band}) with the sum
of the contributions from the new bands $\xi_{\pm}(\mathbf{k})$, and it
would produce an extra term $(2/NV) \sum_{\mathrm{RBZ}} (v^F_\a D+v^D_\a
\e)^2 [f(\xi_{+}) - f(\xi_{-})]/E^3 $ that would be added to the spectral
weight (\ref{sum.T>Tc}). Its contribution to $W(T)$ is negative and of
order $t\sqrt{t/D_0}$, as one can check numerically and estimate
analytically at low doping.  The resulting $W(T)$ is then found to decrease
below $T_{DDW}$, in contrast to the experimental observation and the result
obtained with Eq. \pref{sum.T>Tc}.  Analogously, below $T_c$ the ansatz of
coupling the gauge field to the quasiparticle DDW operators does not
reproduced the expression for the superfluid density $\rho_s(T)$ proposed
in Refs.~\cite{Tewari:2001:PRB,Wang:2001:PRL}, which is derived trough the
current operator (\ref{electric.current.DDW}).

The previous discussion shows that there is not yet an agreement
in the literature about the proper treatment of the transport
properties in the DDW state. However, it is worth noting that the
form of the current operator and of the diamagnetic term used to
evaluate the electromagnetic response kernel ${\cal K}_{\a\a}(q)$
in the DDW state are intimately related. If a Gauge invariant
approximation is used, the response kernel satisfies ${\cal
K}_{\a\a}(\o=0,\bq\ra 0)=0$ above the SC critical temperature
$T_c$ \cite{Scalapino:1993:PRB, Schrieffer}. This means for
example that the diamagnetic contribution $\langle \tau_{\a \a}
\rangle$ to the superfluid density $\rho_s(T)$ cancels the
contribution $\Lambda_{\a \a}(i\Omega_n =0, \mathbf{q} \to 0)$
providing the vanishing of $\rho_s(T)$ for $T > T_c$.  Within the
low-energy model \pref{Hamiltonian} this cancellation holds only
if the diamagnetic term \pref{new.tensor} is considered along with
the current operator \pref{electric.current.DDW}, derived from the
requirement that the continuity equation
(\ref{charge.conservation}) be satisfied.The same result does not
hold by using the mean-field correlation functions defined in Ref.
\cite{Chakravarty:2002:PRL}.

A different approach, which was not investigated here, consists of deriving
a proper gauge-invariant approximation for the response kernel ${\cal
K}_{\a\a}$ by starting from an underlying microscopic model that provides
the basis for the Hamiltonian \pref{Hamiltonian}
\cite{Halperin:1968:SSP,Affleck:1988:PRB,Nersesyan:1989:JLTP,
Schulz:1989:PRB,Varma:1997:PRB,Eremin:1998:JETPL,Cappelluti} and including
the vertex corrections to the mean-field correlation functions. In the case
of SC symmetry breaking, one knows that vertex corrections are singular for
$(\omega,\bq)\ra 0$, satisfying the dispersion relations of the collective
(phase) mode \cite{Enz:book,Schrieffer}.  In the DDW case, where the phase
mode is locked by the commensurability, vertex corrections are always
finite and at zero frequency are related by Ward identities to the ${\bf
k}$ derivative of the self-energy associated with the DDW state, i.e.  to
the term $v_\bk^D$ which appears in the definition
(\ref{electric.current.DDW}) of the current. As a consequence, as observed
in Ref.~\cite{Sharapov:2003:PRB}, the d.c. conductivity, $\sigma(0)$
calculated with the current operator (\ref{electric.current.DDW}) coincides
with the exact results for a general, many-body formulation with nonzero
vertex corrections. This observation suggests that there exists an energy
scale below which the approach followed here, where the sum rules
(\ref{sum.T>Tc}) for $T > T_c$ and (\ref{sum.complete}) for $T < T_c$ were
derived directly from the effective Hamiltonian \pref{Hamiltonian},
describes properly the system behavior. However, since there is no
straightforward extension of the previous arguments for $\sigma(\omega \neq
0)$, it is difficult to determine the value of $\omega_m$ at which the
restoration of the more general sum rules (\ref{weight.band}) and
(\ref{full.rule}) should be observed in the original microscopic model. In
particular, if this cut-off energy resulted to be quite lower than the bare
plasma edge, the comparison with the experimental data presented before
should be reconsidered and referred to the data collected up to frequencies
lower than $10^4$ cm$^{-1}$.  As a consequence, the restricted sum rule
derived here for the mean-field Hamiltonian \pref{Hamiltonian} has yet to
be understood from a more general point of view, within a direct analysis
of a microscopic Hubbard-type model with some short-range interaction that
may result in the formation of DDW state. As mentioned above, this
investigation is rather complicated and cannot be done within the framework
considered here, so we reserve it for a future work.

To conclude, we have demonstrated within an effective model that
circulating currents can act to modify the restricted optical sum rule
in a such way that this acquires the same temperature dependence as
that observed in experiments above $T_c$: the opening of the
corresponding gap produces an {\em increase} in the spectral weight
above $T_c$. Below the SC critical temperature the spectral weight
keeps almost constant, as observed recently in
Ref. \cite{Santander:2004}, but in contrast with other measurements
\cite{Molegraaf:2002:Science}. Since the experimental situation about
the behavior of the spectral weight in the SC state is not settled,
more data are certainly required to definitively establish the
possible compatibility between our findings and the experiments. As
far as our theoretical approach is concerned, we discussed that the
exact range of validity of this result should still be
clarified. Nonetheless, the analysis of the reduced low-energy model
suggests the possibility that the same kind of deviations from the
conventional form of the restricted sum rule could be expected in more
sophisticated microscopic models (see e.g. \cite{Varma:1997:PRB}).

\section{Acknowledgments}

We are grateful to N.~Andrenacci, V.P.~Gusynin, C.~de~Morais~Smith,
B.~Normand, and
P.~Prelov{\v s}ek for helpful discussions.
This work was supported by the research projects
2000-067853.02/1 and 620-62868.00 of the Swiss NSF.

\end{document}